\documentclass[aps,prl,reprint,superscriptaddress]{revtex4-1}
\usepackage{amsmath,amssymb}
\usepackage{bm}
\usepackage{graphicx,color,natbib}
\usepackage{placeins}
\usepackage[dvipsnames]{xcolor}
\usepackage[compact]{titlesec}

\newcommand{\beq}{\begin{equation}}
\newcommand{\eeq}{\end{equation}}
\newcommand{\beqa}{\begin{eqnarray}}
\newcommand{\eeqa}{\end{eqnarray}}

\newcommand{\ds}{\displaystyle}

\newcommand{\pecl}{\operatorname{\mathit{P\kern-.08em e}}}

\titlespacing{\section}{0pt}{*1}{*0}

\begin{document}

\title{Competing Active and Passive Interactions Drive Amoeba-like Living Crystallites and Ordered Bands}

\author{Abraham Mauleon-Amieva}
\affiliation{H.H. Wills Physics Laboratory, Tyndall Avenue, Bristol, BS8 1TL, UK}
\affiliation{School of Chemistry, University of Bristol, Cantock's Close, Bristol, BS8 1TS, UK}
\affiliation{Centre for Nanoscience and Quantum Information, Tyndall Avenue, Bristol, BS8 1FD, UK}
\affiliation{Bristol Centre for Functional Nanomaterials, Tyndall Avenue, Bristol, BS8 1FD, UK}

\author{Majid Mosayebi}
\affiliation{School of Mathematics, University of Bristol, Bristol, BS8 1TW, UK}
\affiliation{BrisSynBio, Life Sciences Building, Tyndall Avenue, Bristol, BS8 1TQ, UK}

\author{James E. Hallett}
\affiliation{H.H. Wills Physics Laboratory, Tyndall Avenue, Bristol, BS8 1TL, UK}
\affiliation{School of Chemistry, University of Bristol, Cantock's Close, Bristol, BS8 1TS, UK}
\affiliation{Centre for Nanoscience and Quantum Information, Tyndall Avenue, Bristol, BS8 1FD, UK}

\author{Francesco Turci}
\affiliation{H.H. Wills Physics Laboratory, Tyndall Avenue, Bristol, BS8 1TL, UK}

\author{Tanniemola B. Liverpool}
\affiliation{School of Mathematics, University of Bristol, Bristol, BS8 1TW, UK}
\affiliation{BrisSynBio, Life Sciences Building, Tyndall Avenue, Bristol, BS8 1TQ, UK}

\author{Jeroen S. van Duijneveldt}
\affiliation{School of Chemistry, University of Bristol, Cantock's Close, Bristol, BS8 1TS, UK}

\author{C. Patrick Royall}
\affiliation{H.H. Wills Physics Laboratory, Tyndall Avenue, Bristol, BS8 1TL, UK}
\affiliation{School of Chemistry, University of Bristol, Cantock's Close, Bristol, BS8 1TS, UK}
\affiliation{Centre for Nanoscience and Quantum Information, Tyndall Avenue, Bristol, BS8 1FD, UK}

\begin{abstract}
Swimmers and self-propelled particles are physical models for the collective behaviour and motility of a wide variety of living systems, such as bacteria colonies, bird flocks and fish schools. Such artificial active materials are amenable to physical models which reveal the microscopic mechanisms underlying the collective behaviour. Here we study colloids in a DC electric field. Our quasi-two-dimensional system of electrically-driven particles exhibits a rich and exotic phase behaviour. At low field strengths, electrohydrodynamic flows lead to self-organisation into crystallites with hexagonal order. Upon self-propulsion of the particles due to Quincke rotation, we find an ordered phase of active matter in which the motile crystallites constantly change shape and collide with one another. At higher field strengths, this ``dissolves'' to an active gas. We parameterise a particulate simulation model which reproduces the experimentally observed phases and, at higher field strengths predicts an activity-driven demixing to band-like structures.

\end{abstract}

\maketitle

\section*{Introduction}
From living organisms to synthetic colloidal particles, active systems display exotic phenomena not attainable by matter at thermal equilibrium \cite{marchetti2013,schweitzer2002,ramaswamy2010,bechinger2016}, such as swarming \cite{ariel2015,narayan2007}, cluster-formation \cite{theurkauff2012,palacci2013} or phase separation in the absence of attractions \cite{cates2015,buttinoni2013,schwarzlinek2012}, banding \cite{chate2008}, and unusual crystallisation behaviour \cite{briand2016}. This is due to continuous energy consumption which occurs in a wide range of systems at very different lengthscales, from the cell cytoskeleton \cite{julicher2007,sanchez2012}, tissues \cite{park2014} and bacterial colonies  \cite{pedley1992,zhang2010, lushi2014,petroff2015} to larger scales such as insect swarms \cite{sinhuber2016}, fish schools \cite{katz2011} and bird flocks \cite{cavagna2010}. Artificial active materials, composed of microswimmers, active colloids or vibrating granular particles \cite{zhang2010,volpe2011,theurkauff2012,bechinger2016,briand2016}, or even synthetically modified living systems such as bacteria \cite{schwarzlinek2012}, provide a suitable testing ground where the behaviour of active matter may be carefully probed to extract the new physical principles of this emergent class of matter.

While simple models of active particles capture some of the complex behaviour observed experimentally, for example collective motion \cite{vicsek1995,gregoire2004,redner2013,fodor2016,zottl2014}, the link between experiment and theory in active matter is often rather qualitative. As a result, a comprehensive understanding of how and which microscopic mechanisms lead to the emergence of complex structures in experimental active systems remains elusive. Here we implement a theoretical description which is able to predict the behaviour observed in experiments. In particular, we parameterise our experimental system at the microscopic level of the interacting particles. We combine this model with particle-resolved studies of so-called Quincke rollers, active colloids which exhibit swarming and flocking \cite{bricard2013,bricard2015}.

At low--to--moderate motility, we reveal the importance of competing passive interactions (long-ranged attractions) driving crystallisation and activity which leads to melting--like and evaporation--like behaviour.  At high motility, the role of passive and active interactions is \emph{reversed}: activity drives demixing resulting in a banding phase, whose ordered local structures result from the repulsive core of the particles. This importance of competition between passive and active interactions is reminiscent of well-known systems such as amphiphiles, block copolymers and mixtures of charged colloids and non-absorbing polymer where competing interactions lead to modulated phases such as lamellae \cite{andelman1995,ciach2008}, whose structures indeed resemble some we find here. Our approach shows how one may build bottom-up designs of particulate active matter with precisely controllable macroscopic behaviour.

\begin{figure*}
\centering
\includegraphics[width=0.99\textwidth]{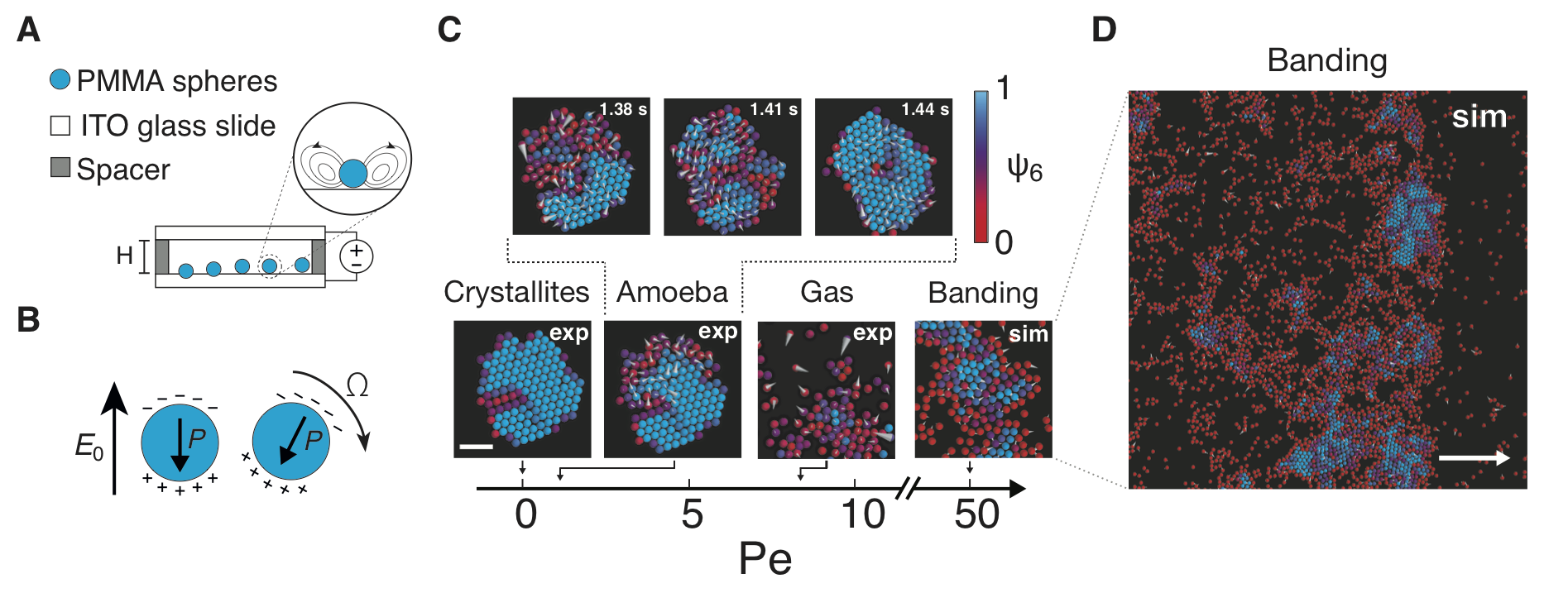}
\label{figPhaseSetUp}

\caption{\textbf{Phase diagram of Quincke Rollers as a function of activity.}
\textbf{A} \textbf{Experimental set-up.} Colloidal particles are suspended and confined within a sample cell made of conductive glass slides. The induced electrohydrodynamic flow is represented with dashed lines in the amplified illustration.  This flow field leads to long-ranged electrohydrodynamic interactions between the particles \cite{nadal2002}.
\textbf{B.} \textbf{Schematic representation of mechanism of Quincke rolling.} The charge distribution around the sphere forms a dipole oriented inversely to the field direction, and any fluctuation in the dipole orientation leads to particle rotation
with a constant angular speed $\Omega$. The field-dependent activity is translated to the P\'eclet number, as described in the text.
\textbf{C. Phase behaviour as a function of P\'eclet number.} In the low activity regime, electrohydrodynamic interactions due to flow fields (see inset in Fig. \ref{figPhaseSetUp}\textbf{A}) result in (passive) crystallite formation. On increasing the activity, e.g. $\pecl$ = 1.5 ($E_0 = 19.4 E_{Q}$, with $E_{Q} \approx 6.2\times10^4\,\rm{V}\cdot m^{-1}$), the particles self-propel sufficiently that the dynamics change markedly (see upper sequence), and the active crystals split and \textbf{coalesce} with one another. With a further increase of activity, the crystallites melt and we find an active gas. In the regime accessible to our simulations, we find banding at a somewhat higher area fraction of $\phi=0.08$. Scale bar is 10 $\mu$m. Colorbar indicates the local hexagonal order of coloured particles, $\psi_6$.
\textbf{D. Banding.} Zoomed out rendering of the band formation in simulations. White arrow indicates the direction of propagation.
}
\end{figure*}

In the Quincke rollers we study, the application of a uniform DC electric field above a critical field strength $E_Q$ induces the directed motion of spherical colloids by coupling their translation and diffusion near a surface \cite{jakli2008,bricard2013}. In the absence of a field, the particles behave as conventional passive Brownian colloids. At low field strengths, while remaining non-motile, particles agglomerate into crystals due to to long-ranged attractive interactions which arise from electro-osmotic flows (Fig.\,\ref{figPhaseSetUp}\textbf{A}.) \cite{yeh2000,ristenpart2004,zhang2004}. Above the critical field strength $E_Q$, the particles undergo Quincke rotation \cite{quincke1896,pannacci2007,das2013} and become motile (Fig.\,\ref{figPhaseSetUp}\textbf{B}) so that the electro-osmotically generated crystallites transition into a highly mobile active state reminiscent of amoebae (see Supplementary Movies 1-3, available online in \textit{https://abrhmma0.wixsite.com/website}). Unlike ``living crystals'' \cite{palacci2013}, they are motile and characterised by a highly dynamic outer surface. These crystallites then dissolve into an isotropic active gas as we increase the field strength. Finally at very high field strengths our simulations predict that the system undergoes banding which involves local ordering. We investigate the rich structural and dynamical properties of our system using a range of static and dynamic order parameters and in particular consider the coalescence and division of our amoeba-like motile crystals.

\vspace{1cm}
\section*{Results}
\subsection*{Experiments}
A more complete description of the experimental setup shown schematically in Fig.\,\ref{figPhaseSetUp}\textbf{A}. is included in the Methods section. Briefly, we use a suspension of colloidal particles of diameter $\sigma=2.92$\ $\mu$m in a non-aqueous ionic solution. Sedimentation results in a quasi 2D system with area fraction $\phi_\mathrm{exp}\approx0.05$. The sample cell is made from two indium-tin-oxide (ITO) coated glass slides separated with UV-cured resin containing spacer beads to allow the application of the electric field. The samples are imaged with bright-field optical microscopy for particle tracking. A uniform field $E_0$ is applied perpendicular to the two slides as represented in Fig.\,\ref{figPhaseSetUp}\textbf{A}, in order to enable self-propulsion due to Quincke rotation. We then translate the resultant field-dependent activity to dimensionless P\'eclet numbers and characterise the static and dynamic behaviour of the system. Throughout, we use the Brownian time for a colloid to diffuse its own radius in 2D, $\tau=\sigma^2/D_t \approx 9$s, as the unit of time, where $D_t$ is the translational diffusion constant.

\subsection*{Simulations}
The Quincke rollers are subject to forces and torques due to excluded volume repulsions, as well as self-propulsion, alignment and attractions generated by the electro-hydrodynamic interactions of the particles with their environment and each other \cite{nadal2002,bricard2013}. They can be modelled as \emph{active Brownian particles} with an additional {\em active aligning torque}, whose active/passive forces and torques can be quantitatively specified. We implement Brownian dynamics simulations, with the following equations of motion for positions and orientations ${\mathbf{r}}_i,\theta_i$.
\begin{eqnarray}
\dot{\mathbf{r}}_i &=&\frac{D_t}{k_BT} [ {\mathbf{F}}_i + f^p \hat{\mathbf{P}}_i ] + {\sqrt{2D_{t}}} \bm{\xi}_i^{t}, \\
\dot{\theta}_i       &=& \frac{D_r}{k_BT} {\mathcal{T}}_i + \sqrt{2D_r} \xi_i^r,
\label{eqABP}
\end{eqnarray}
where  $\mathbf{F}_i$ is the interparticle force on the $i$th roller, $f^p$ is the magnitude of the active force,
$\hat{\mathbf{P}}_i = ( \cos \theta_i, \sin \theta_i )$ is the direction of motion of the $i$th roller,
${\mathcal{T}}_i$ is the torque on the $i$th roller which incorporates alignment terms, and $\xi_i^{t,r}$ is a Gaussian white noise of zero mean and unit variance.
$D_r$ is the rotational diffusion constant. The direct interactions $\mathbf{F}_i$ include a ``hard'' core and long-range attraction, the latter to model the electrohydrodynamic contribution. Further details of the model and the simulation parameters, and the procedure by which the parameters were mapped to the experiment may be found in the Methods and Supplementary Materials (SM).

\subsection*{Determining the P\'eclet number}
Before moving to the discussion of our results, we first describe our mapping of field strength to P\'eclet number between experiment and simulation. We obtain the bare translational diffusion coefficient of the passive system $D_{t}$ measured at equilibrium. Particle velocity $\upsilon$, and the characteristic time scale for the rotational diffusion $\tau_{r} = D_{r}^{-1}$ for a dilute sample with with area fraction $\phi \approx 0.001$ are obtained from the fitting to the mean square displacement (MSD) of active particles in the dilute (gas) regime,

\begin{equation}\label{eqMsd}
\langle \Delta r^{2} (t) \rangle = 4D_{t}t + \frac{\upsilon^{2}\tau_{r}^{2}}{3}\Bigg[\frac{2t}{\tau_{r}} + \exp\Bigg(\frac{-2t}{\tau_{r}}\Bigg) -1 \Bigg].
\end{equation}
\noindent
To extract the parameters of Eq. \ref{eqMsd} from the experiments we consider a series of similarly dilute samples. We estimate the dimensionless P\'eclet number as  $\pecl=
3\upsilon \tau_{r}/\sigma$, for each measured velocity in the different states obtained in the experiment. The P\'eclet number is significantly enhanced by Quincke rotation. However, since this is related to the threshold field strength $E_Q$ where Quincke rotation is initiated, we find that for low field strengths, $\pecl$ is small and only weakly dependent on the field, $E_0 \ll E_Q, \; \pecl \sim 0$. Once the particles become motile, for our system the two appear to be roughly linearly coupled ($E_0 > E_Q, \;  \pecl\sim E_0$, see SM Fig.\,S1).

We now present our main findings. First we consider the phase behaviour of the system as a function of the activity, represented by the P\'eclet number which we obtain from measuring particle mobility. At zero field strength ($\pecl$ = 0), we obtain Brownian hard discs which form a dilute fluid with area fraction $\phi\approx0.05$. Upon increasing the field strength, the system exhibits a novel phase behaviour owing to a coupling between non-equilibrium electrohydrodynamic interactions due to solvent flow and electrically induced activity (Quincke rotation).

\subsection*{Crystallisation}

Particle condensation to form crystallites emerges at low field strength, e.g.  $\pecl \approx 1\times 10^{-4}$ ($E_0=9.9E_{Q}$). This is due to the long-ranged electrohydrodynamic interactions (Fig.\,\ref{figPhaseSetUp}\textbf{A}) \cite{nadal2002}. In our experiments, colloids
act as dielectric regions perturbing the electric charge distribution, therefore inducing a flow of ions with a component tangential to the substrate \cite{nadal2002}. In the vicinity of such an electro-osmotic flow, the particles experience transverse motion leading to the formation of crystallites (Fig.\,\ref{figPhaseSetUp}\textbf{A}).

\subsection*{Activity-induced phase transitions}

Upon increasing the field strength, we can exploit the Quincke mechanism that triggers spontaneous rotation (Fig.\,\ref{figPhaseSetUp}\textbf{B}) to study the behaviour of self-propelled rollers. For this to occur, the viscous torque acting on the particle must be overcome, hence the field needs to be sufficient to initiate rolling ($>E_Q$).  When increasing the activity above $\pecl$ = 1.5 ($E_0 = 19.4E_{Q}$), we observe crystallite motility, coalescence and splitting, yet the local hexagonal symmetry remains, as can be seen in certain bacteria colonies \cite{petroff2015} and chiral swimmers \cite{shen2019}. We term this an ``amoeba phase'', since the motility leads the aggregate to constantly reshape in a fashion reminiscent of the motion of amoebae, as shown by the time sequence in Fig. \ref{figPhaseSetUp}\textbf{c} (also see Supplementary Movies 1-3).

On further increasing the field to $E_0 = 29.8E_{Q}$ ($\pecl$ = 7.8), Quincke rotation triggers breakdown of the active crystallites into an ``active gas'' of colloidal rollers undergoing displacement in random directions, Fig.\,\ref{figPhaseSetUp}\textbf{C}. Previously, it was shown experimentally that the increase in area fraction results in homogeneous polar phases and vortices \cite{bricard2013,bricard2015}. At higher field strength, our simulations predict that around $\pecl \gtrsim 35$, the system exhibits a non-equilibrium phase transition to a banded state. These bands form perpendicular to the direction of particle motion (which self-organizes into a strongly preferred direction). (see Supplementary Movies 4 and 5). This is reminiscent of banding observed in earlier experiments \cite{bricard2013}, but here the area fraction is very much higher, leading to local hexagonal order. In our simulations, we see one band in the box. We leave the analysis of whether this is
activity-driven micro-phase separation, or full demixing for a later finite-size scaling analysis. This local hexagonal order within the bands contrasts with the unstructured bands seen in the Vicsek model \cite{chate2008}.

\begin{figure*}
\centering
\includegraphics[width=0.99\textwidth]{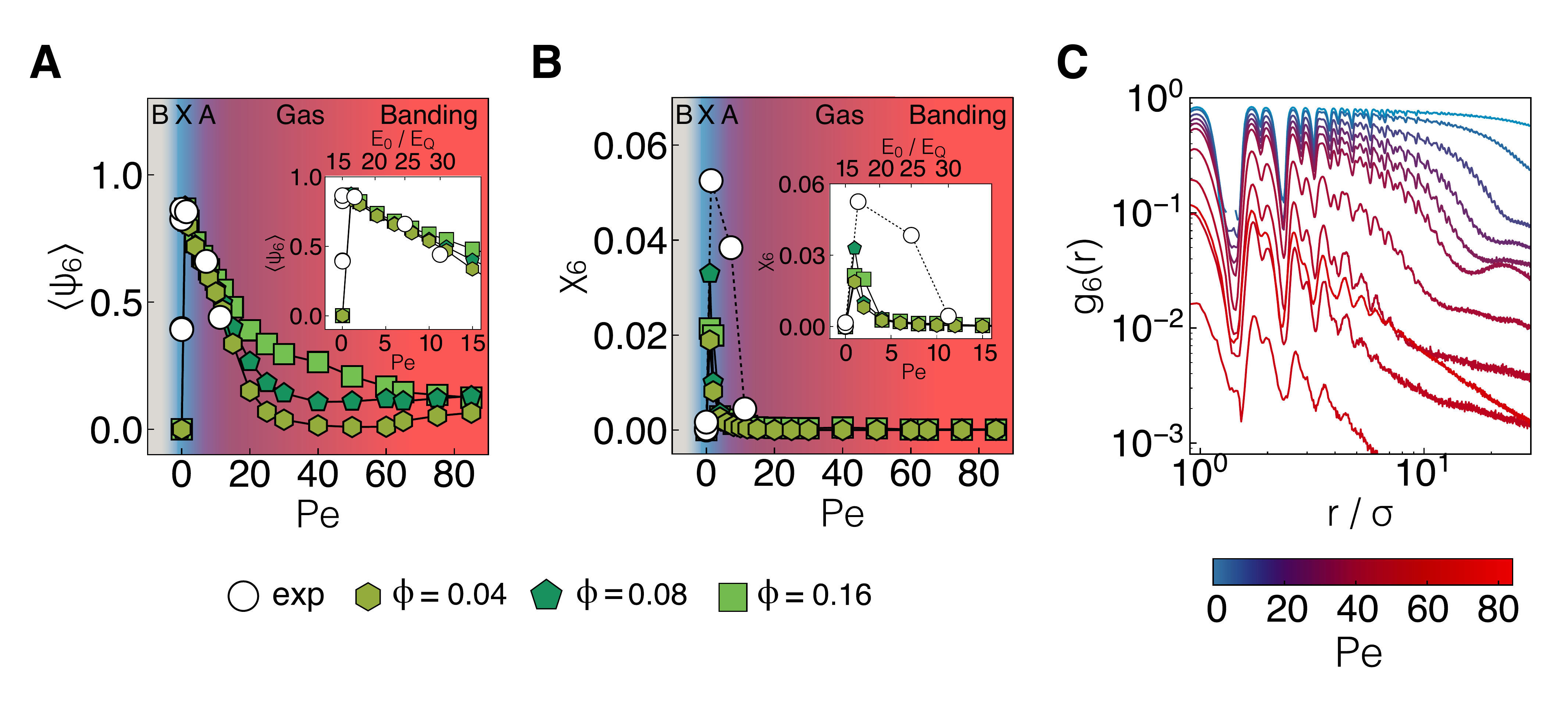}
\caption{\textbf{Changes in local structure as a function of field strength for Quincke rollers.}
\textbf{A}. Local order determined with the bond-orientational parameter $\psi_6$ upon increasing field strength.
\textbf{B}. Fluctuations of the bond-orientational parameter $\chi_6$ upon increasing field strength as defined in the text.
Dashed line indicates the onset of motility, which distinguishes crystalline and amoeba states.
Insets in \textbf{A} and \textbf{B} show data plotted with respect to the applied field strength.
\textbf{C}. Orientational correlation functions $g_{6}(r)$ for $\pecl$ as indicated in the color bar. Data obtained from simulations with $\phi = 0.08$.
In \textbf{A} and \textbf{B} and in subsequent figures, the phases are denoted as B - Brownian fluid, X - crystallites, A - amoebae, (active) gas and banding. Insets in \textbf{A} and \textbf{B} are zoom in sections indicating $E_{0}$ values.
}
\label{figStructure}
\end{figure*}

\subsection*{Analysis of the Local Structure}

Having qualitatively introduced the behaviour we encounter in our system in Fig.\,\ref{figPhaseSetUp}, we now proceed to consider the phase transitions in more detail. In order to determine the nature of the transitions we require suitable order parameters. We first consider the structural properties of the phases we encounter --- passive fluid, passive crystal, active crystallite (``amoebae''), active gas and bands. Given the richness of the phase behaviour, 
it is unlikely that one single order parameter will prove sufficient, and we find this to be the case. We begin with the 2D bond-orientational order parameter, $ \psi_6 =(1/N)\sum_{i=1}^{N}|\psi_6^i|$. Perfect hexagonal ordering is indicated by $\psi_{6} =1$, whereas a completely disordered configuration gives $\psi_{6} =0$. See Methods for more details of $\psi_6$.

In Fig. \ref{figStructure}\textbf{A}, we plot the average $\psi_{6}$ as a function of $\pecl$ for both experiment and simulation, under our mapping of $\pecl$ between the two and in the inset we show the data with respect to the applied electric field. In the SM, we show $\psi_6$ with respect to the electric field strength $E_0$. We emphasise that, given the simplicity of our model, and of our mapping, the agreement between experiment and simulation is remarkable in the experimentally accessible regime ($\pecl$ $\lesssim 8$). We find almost no ordering for the passive Brownian system (at $E_0=0$ or $\pecl$ $=0$). With a slight increase in the field strength to $E_0=9.9E_{Q}$, we observe a rapid rise in $\psi_{6}$ to $\approx 0.9$ that corresponds to the crystallisation transition driven by the electrohydrodynamic interactions. In this regime, the system is composed of many crystallites that barely move. It is possible that there may be a condensed liquid (or hexatic) phase \cite{bernard2011}, although this is not apparent in our data, and the transition appears first-order within the field strengths we have sampled. We believe this to be similar to equilibrium 2D attractive systems undergoing crystallisation and move on to consider the activity-driven transitions.

Increasing the activity further into the amoeba phase, $\psi_{6}$ starts to decrease. However, $\psi_6$ remains significantly above zero indicating the amoeba clusters are crystal-like. While this state is far-from-equilibrium, the $\psi_6$ value exhibits temporal fluctuations consistent with a steady state (Supplementary Fig. S2). We infer that to distinguish the (passive) crystallites from the amoebae, some kind of dynamic order parameter may prove suitable, and return to this below. At larger $\pecl$ ($10\lesssim\text{$\pecl$}\lesssim40$), the value of $\psi_6$, drops markedly, as the amoebae ``dissolve'' into the active gas, apparently in a continuous fashion. Finally at very high $\pecl$ ($\pecl\gtrsim40$), upon the emergence of banding, a form of motility-driven phase separation, the value of  $\psi_{6}$ again shows signs of increase for $\phi < 0.16$.

To gain further insight into these transitions, in Fig.\,\ref{figStructure}\textbf{B} we plot the fluctuations in the hexatic bond-orientation order parameter which we take as $\chi_6 = \langle \psi_6^2 \rangle - \langle\psi_6\rangle^2$ where the average is over different snapshots. Further details are provided in the Methods. At low P\'eclet numbers, we see good agreement between the experiment and simulation, but when the motility is higher, the simulations decay towards the active gas faster than the experiments. However, we find no enhancement in $\chi_6$ around the amoeba-gas phase boundaries, indicating that the transition is a cross-over rather than a first-order-like transition between different phases.

To quantify the spatial correlations in $\psi_6$, in Fig.\,\ref{figStructure}\textbf{C}, we plot $g_6(r)$ defined as,
\begin{equation}
g_6(r=|\mathbf{r}_i - \mathbf{r}_j|) = \langle   {\psi_6^i}^*  {\psi_6^j}  \rangle
\end{equation}
where $\psi_6^i$ is the (complex) value of the hexatic bond-orientation order parameter for particle $i$ at position $\mathbf{r}_i$. At low $\pecl$, we observe long-ranged orientational correlations in the crystal and amoeba regimes. Such correlations are significantly shorter-ranged for the active gas. Interestingly, for the largest $\pecl$ in the banding regime, we find that the bond-orientational order parameter is correlated over a larger domain than in the gas regime. Therefore, formation of the bands not only increases 
$\psi_6$, but also enhances its spatial correlations.

\begin{figure*}
\centering
\includegraphics[width=0.9\textwidth]{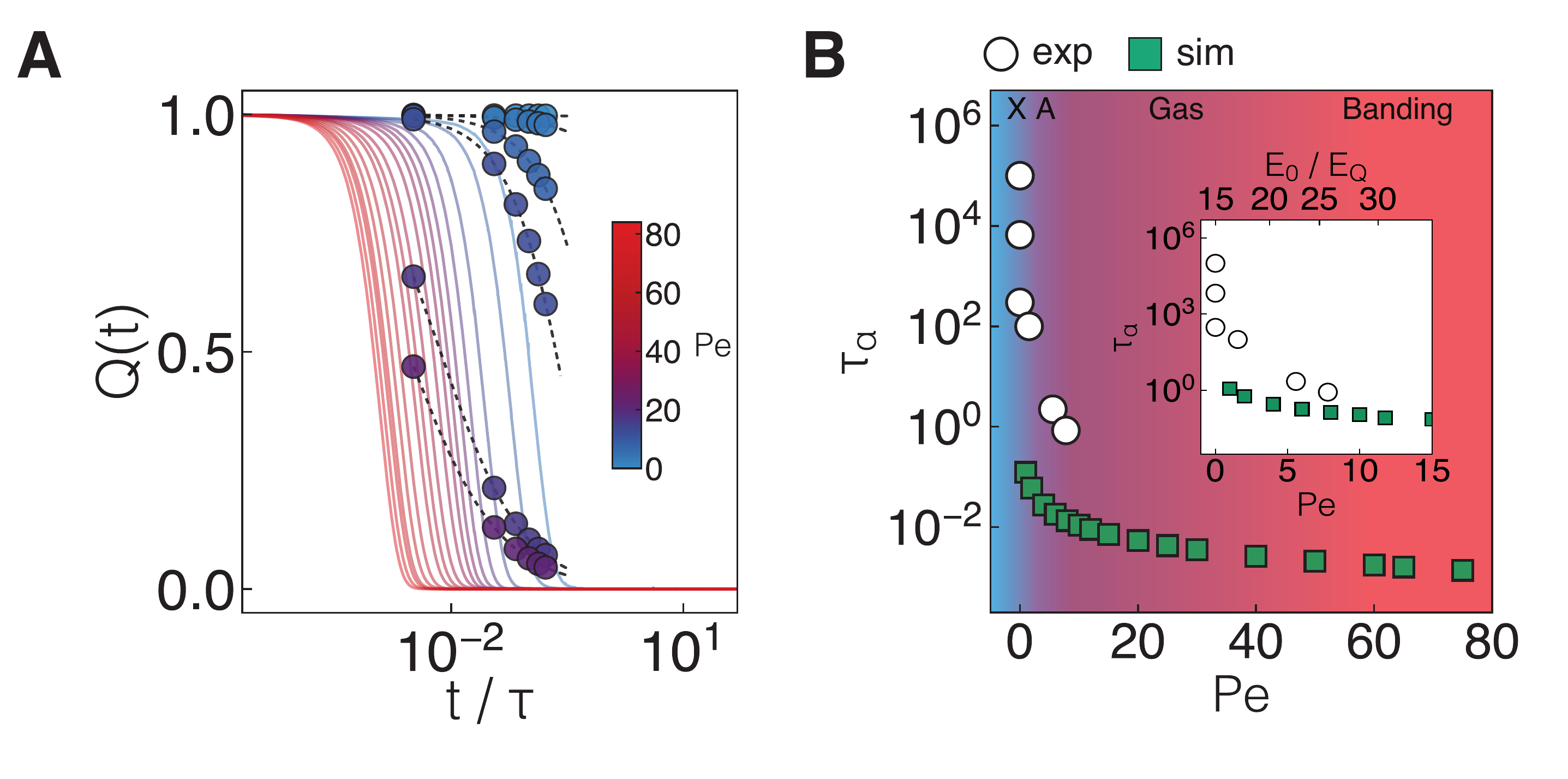}
\caption{\textbf{Dynamics of the Quincke Rollers across various phases.}
\textbf{A.} Dynamical overlap function $Q(t)$, using Eq. \ref{qt}. Coloured circles are experimental data, and solid lines are obtained from simulations.
Dashed lines are stretched exponential fits as described in the text. Color bar indicates the correspondent $\pecl$ for each line. Data are scaled by the Brownian time $\tau$.
\textbf{B.} Relaxation time $\tau_{\alpha}$ from stretched-exponential fitting to symbols and lines. Inset shows the relaxation time plotted as a function of the field strength in the regime accessible to the experiments.}
\label{figDynamics}
\end{figure*}

\subsection*{Dynamical Analysis}

In our analysis of the local structure in Fig. \ref{figStructure}, we noted that some kind of dynamical order parameter would be appropriate to distinguish the crystallites from the amoebae. In Fig. \ref{figDynamics}, we use such an order parameter to perform this analysis, the overlap \cite{briand2016},

\begin{equation}\label{qt}
Q(t) = \Bigg\langle \frac{1}{N}\sum^{N}_{i=1} \exp  - \Bigg(
\frac{\big[\textbf{r}_{i}(t'+t) - \textbf{r}_{i}(t')\big]^2}{a^2} \Bigg)\Bigg\rangle_{t'},
\end{equation}

\noindent
which we evaluate at $a=\sigma$. We fit the resulting dynamic correlation functions with a stretched exponential form, $Q(t) = \exp[-(t/\tau_\alpha)^b]$, as shown in Fig. \ref{figDynamics}\textbf{A} to determine a timescale for relaxation in our system, $\tau_\alpha$. We plot this timescale against the P\'eclet number in Fig.\,\ref{figDynamics}\textbf{B}.

Most striking in the crystal-amoeba transition is the massive drop in relaxation time, Fig. \ref{figDynamics}\textbf{B}: at a total of \emph{six decades}, this is a very substantial change dynamical change for particle-resolved studies of colloids, active or passive \cite{ivlev}.  The crystallites are effectively solids, while the amoebae exhibit timescales of colloidal liquids, even though their structure is crystalline. Despite this precipitous drop in relaxation time, we find that the transition from crystallites to amoeba is apparently continuous in nature. We thus conclude that the crystallite-amoebae and amoeba-active gas transitions we have found are both continuous, at least insofar as we can detect.

\begin{figure*}
\centering
\includegraphics[width=\textwidth]{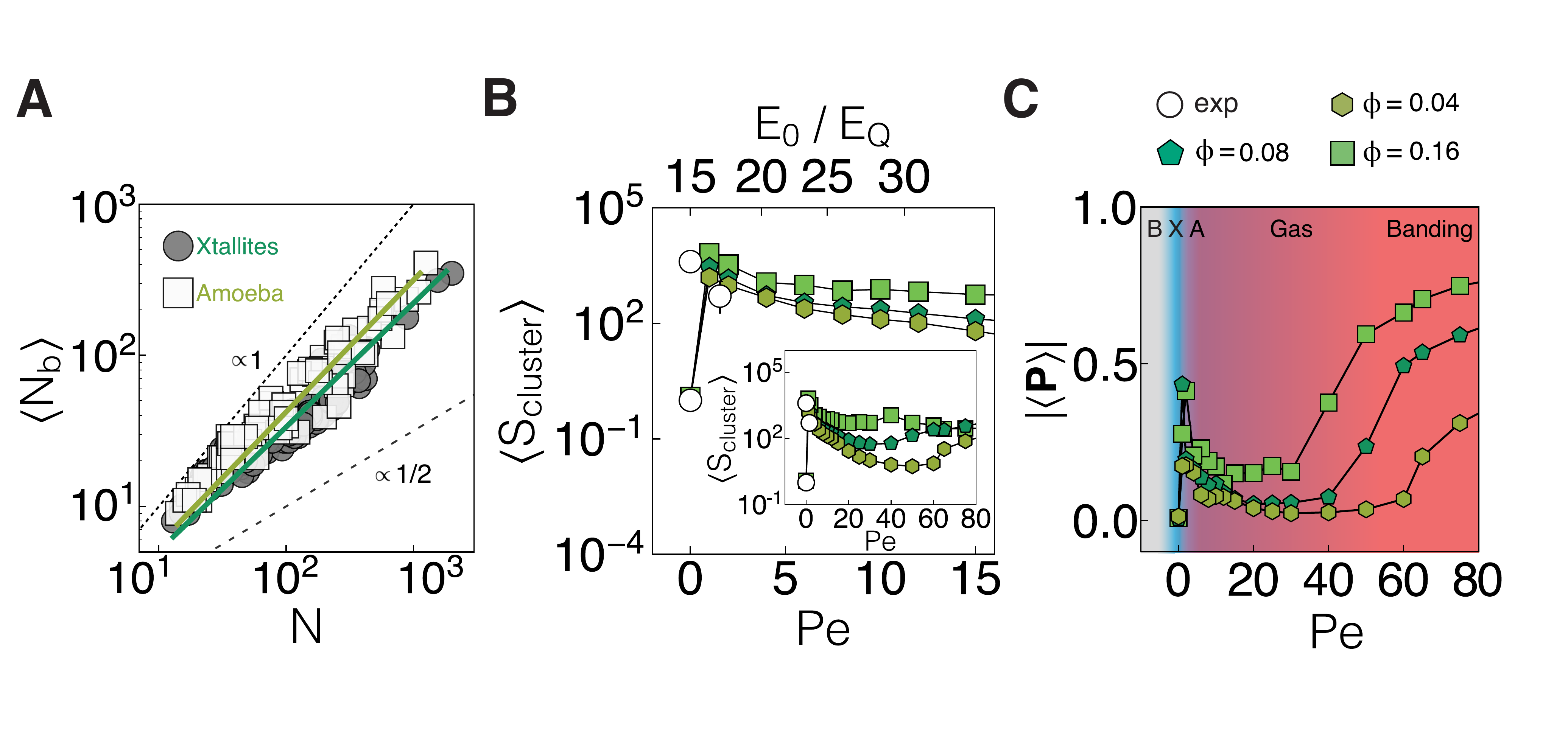}
\caption{\textbf{Characteristics of the clusters formed by the Quincke rollers.}
\textbf{A}. Interfacial broadening in crystallites and amoebae. This is averaged over all the particles sampled in the experiment at each state point.
\textbf{B}. Size of clusters as a function of activity. Inset is a zoom in 
to indicate the field values.
\textbf{C}. Dipolar orientations as function of field strength.}
\label{figClusterDipole}
\end{figure*}

\subsection*{Characteristics of the Active and Passive Crystallites}

In Fig.\,\ref{figStructure}, our $\psi_6$ bond-orientational order parameter gave somewhat limited insight as to the nature of the crystallite-amoeba transition, as both exhibit hexagonal local symmetry. Therefore we now seek other structural measures. In Fig.\,\ref{figClusterDipole}\textbf{A}, we consider the fractal nature of the crystallites formed, be they passive or active. The images in Fig. \ref{figPhaseSetUp} and Supplementary Movies 1-3 hint that, if the amoebae are rather mobile, then the ``interface'' between the ``amoeba'' and its surroundings may be broadened, leading to more particles at the interface and a lower fractal dimension. Specifically, for each cluster we count the number of particles identified on the boundary $N_b$. In the case of compact clusters, this should scale with the number of particles in the cluster as $N^\nu$ with $\nu=1/2$. Figure \ref{figClusterDipole}\textbf{A} shows the different nature of the passive and active clusters with $\nu \approx$ 3/4 and 5/6 respectively, indicating a 
rougher boundary for the active clusters.

Figure \ref{figClusterDipole}\textbf{B} shows how the mean cluster size varies in different regimes. The system is composed of a few large clusters at very low $\pecl$. Upon increasing the activity, those big clusters break up to smaller ones until in the gas regime where the system is dominated by monomers. Consistent with our previous observations, the cluster size is non-monotonic, with an increase of the mean cluster size with the formation of the bands at large $\pecl$. Note that in the regime where our simulations indicate banding, finite size effects in the simulations (which have $N=10000$ particles) may influence the cluster size somewhat as the bands span the simulation box. The same holds for the passive crystals at low field strength.

\subsection*{Nature of the Transitions at Higher Activity: Amoeba to Active Gas and Active Gas to Ordered Bands}
In addition to the transitions we have already discussed, we encounter more at higher field strength. Firstly, the amoebae ``dissolve'' to form an ``active gas''. At the densities we consider, this transition is characterised by a substantial -- but continuous -- drop in the $\psi_6$ bond-orientational order parameter (Fig. \ref{figStructure}\textbf{A}) consistent with our discussion of the continuous change in dynamics above.

At higher field strengths, we encounter banding, strong density fluctuations perpendicular to the preferred direction of travel. Interestingly, these bands exhibit some degree of local order, as the value of the bond-orientational order parameter $\langle \psi_6 \rangle \approx 0.2$. While far from indicating full hexagonal order ($\langle \psi_6 \rangle = 1$), this is nevertheless significantly larger than zero. Furthermore, as we can see in Fig. \ref{figPhaseSetUp}\textbf{C}, some particles are in a very high state of crystalline order (appearing blue), although most are not. Rather striking, in the case of the transition to the banded phase is the alignment between the dipoles of the Quincke rollers, which defines the direction of rotation (see SM for details). In Fig.  \ref{figClusterDipole}\textbf{C}, we see a very strong increase in the alignment upon banding, suggesting that this is a suitable order parameter in this case.

\begin{figure*}
\centering
\includegraphics[width=0.75\textwidth]{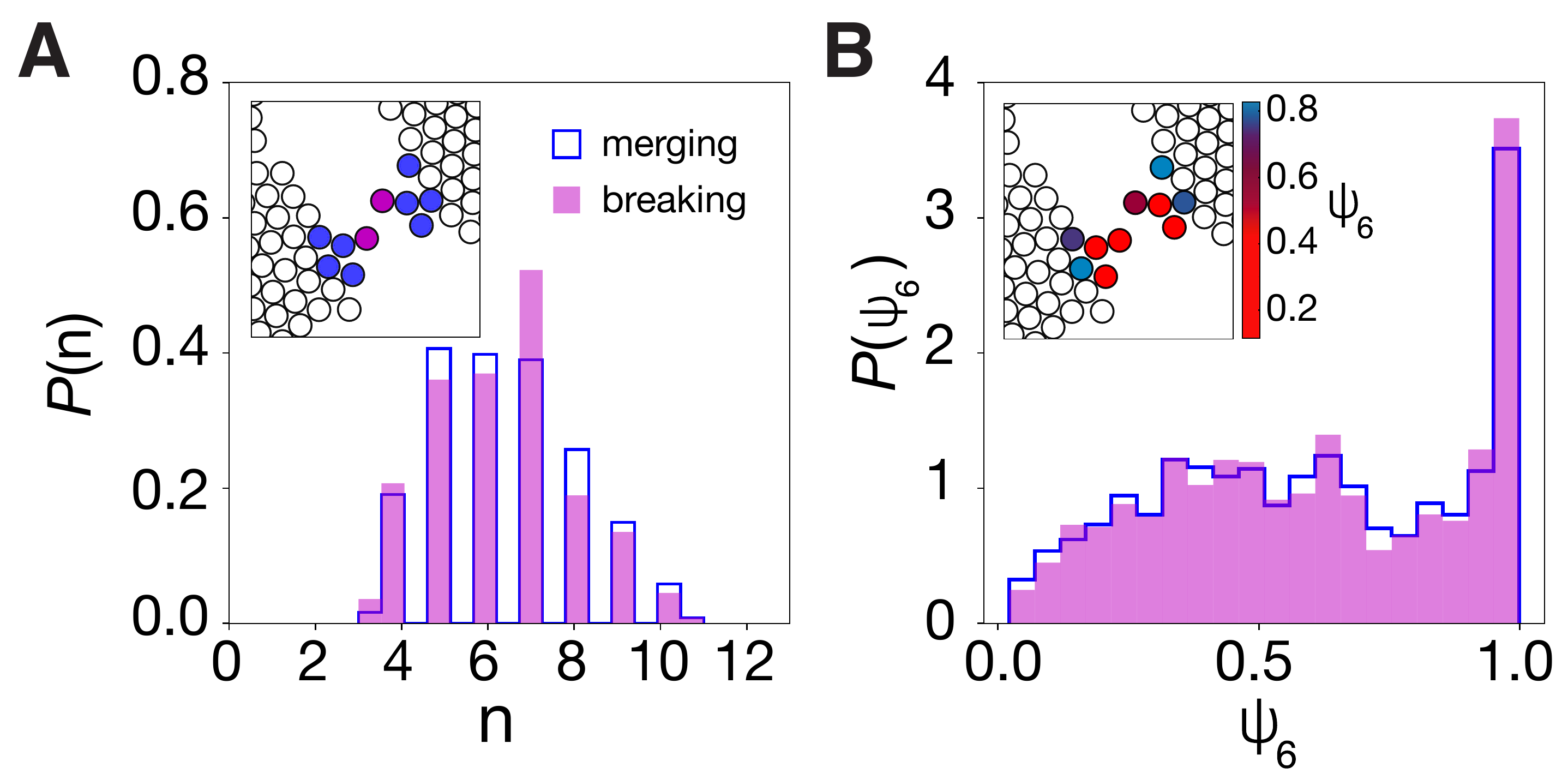}
\caption{\textbf{``Amoeba'' coalescence and splitting.}  \textbf{A.} Distribution of number of shell (blue) particles for all amoebae
either splitting or coalescing. Inset: Snapshot prior to a coalescence event between two clusters. Purple particles represent the nearest boundary particles between clusters, and blue particles are the first shell of particles around purple particles. \textbf{B.} Distribution of the local $\psi_{6}$ for the same breaking and merging events. Inset: Same snapshot now showing the local hexagonal order for the same particles.}
\label{figAmoebaMerging}
\end{figure*}

\subsection*{Amoeba Coalescence and Splitting}
As Supplementary Movies 1 and 3 show, the amoebae are highly dynamic. Here we analyse their coalescence and splitting.  It is intriguing to compare it with coalescence and splitting (or break-off) of liquid droplets \cite{eggers1999,eggers2008}. However there are several key differences between the two situations. Both coalescence and break-off of liquid drops in passive systems are macroscopic driven processes that are clearly {\em not} in a steady state \cite{eggers1999,eggers2008}, while here the amoeba phase is apparently a non-equilibrium steady state. Furthermore here the clusters are locally ordered. We consider the behaviour at the particle scale and emphasise the 2d nature of the system. Nevertheless one may ask if  insights about the steady-state behaviour of our active system can be obtained by considering the behaviour of a driven passive system that is not in a steady state. It is noteworthy that in passive systems there is a strong asymmetry between liquid drop coalescence and break-off   -- break-off dynamics is very distinct from that of coalescence, reflecting the fact that end-points of the two processes are rather different. In contrast, thermodynamic equilibrium is a steady state with vanishing currents and hence one expects symmetry  between coalescence and break-off at equilibrium. However it is not at all obvious if there would be a symmetry between the dynamics of coalescence and break-off in the amoeba phase because it is a non-equilibrium steady-state. That is to say, while of course the microscopic equations of motion in this non-equilibrium system are not expected to exhibit time-reversal symmetry, we enquire whether this is apparent at the level of the coalescence and splitting behaviour in the Amoeba phase. In particular, we consider whether we can distinguish the pathways by which coalescence and splitting occurs.

We analyse the coalescence and break-off events in the amoeba phase as follows. First distinct ``amoebae'' are identified (See SM). To investigate the morphological changes, we account for the change in number of particles per ``amoeba'' and measure the nearest distance between boundary particles in different clusters. This yields the particles which first form the link between two ``amoebae'' in the case of coalescence and those where the last point of contact remains in the case of splitting (Fig. \ref{figAmoebaMerging}\textbf{A}, inset).

Having the pair of particles involved in coalescence and splitting, we identify the number of particles within a distance of 1.1 $\sigma$ and analyse the local $\psi_{6}$. In Fig. \ref{figAmoebaMerging}, we plot the resulting distributions of the number of particles $n$ in the neighbourhood and the local $\psi_{6}$ with time running forwards and backwards. Since the distributions appear rather similar within our statistics, we infer that our analysis does not reveal any breaking of time-reversal symmetry, consistent with recent work with active Janus colloids which considered aggregation and fragmentation rates \cite{ginot2018}. Thus this non-equilibrium steady state is fundamrentally different to the highly asymmetric case of droplet coalescence and break-off in driven passive liquids \cite{eggers1999,eggers2008}.

\section*{Discussion}

In conclusion, we have shown that the Quincke roller system exhibits a rich and complex phase behaviour. We have also developed a minimal model of the experimental system which captures all the phenomena observed in experiment, and in fact predicts further phase transitions at activities beyond the experimental regime. We find transitions between passive fluid, crystal, amoeba-like active crystallites, active gas and an ordered banding phase. We have used a variety of static and dynamic order parameters to probe the nature of these transitions, and find that they are continuous in nature except the (passive) fluid-crystal transition which is consistent with first-order. We have analysed coalescence and splitting events in the ``amoeba'' phase, and find that it does not show significant deviations from time-reversal symmetry.

For our simulation model, we have quantitatively parameterised the components of the Quincke roller system by treating the electrohydrodynamic attraction with a long-ranged potential, ``hard'' core,  active force and electrohydrodynamic alignment terms. Remarkably, when we rescale our results to compare the same P\'eclet numbers in experiments and simulations, we obtain a quantitative agreement between the two. With our model, we have revealed that a key ingredient of the phase behaviour is the interplay between \emph{active} and \emph{passive} interactions. At low mobility, (electrohydrodynamic) passive attractions lead to condensation into crystals, which melt with an increase in activity. Conversely, at high activity, the situation is reversed: activity leads to
banding where the (passive) repulsive core controls the local structure. Our work opens the way to using simple, intuitive minimal models which correctly capture the competition between active and passive interactions to describe, \emph{quantitatively}, the macroscopic physical behaviour of complex active systems which are far-from-equilibrium.

\section*{Materials and Methods}
\renewcommand{\thefigure}{S\arabic{figure}}

\subsection*{Experimental setup}
Our experimental model consists of poly(methyl methacrylate) (PMMA) spheres of diameter, $\sigma=2.92\,\mu\mathrm{m}$ determined with SEM. These are suspended in a 5 mM solution of dioctyl sulfosuccinate sodium (AOT) in hexadecane. Imaging and DC field application take place in sample cells made of two indium tin oxide (ITO)-coated glass slides (Diamond Coatings, BO-X-20, 100 nm thick). A separating layer of $H=16.2\,\mu\mathrm{m}$ between the slides is made using larger polymer beads and UV-curable adhesive (Norland 81). The uniform field is applied by connecting the slide to a power supply (Elektro Automatik, PS-2384-05B). Image sequences 
are obtained using brightfield microscopy (Leica DMI 3000B) with a 20x objective and  a frame rate of 100 fps. Individual colloids are identified and particle co-ordinates are tracked using standard methods \cite{crocker1996}.

\subsection*{Determination of the critical strength}
We follow the description of Lemaire and co-workers \cite{pannacci2007,pannacci2009} to estimate the critical field strength, $E_{Q}$. The spontaneous rotation of particles, known as Quincke rotation, strongly depends on the charge distribution at the particle-liquid interface and the respective charge relaxation times, given by $\tau_{p, l} = \epsilon_{p, l}/s_{p, l}$, where $\epsilon_{p,l}$ and $s_{p,l}$ are the dielectric constant and conductivity of the particle and the liquid respectively. In the case of having $\tau_{l} > \tau_{p}$, the induced dipole $\mathbf{P}_\mathrm{exp}$ is stable with respect of field direction. On the other hand with $\tau_{p} > \tau_{l}$, $\mathbf{P}_\mathrm{exp}$ is unstable with respect to the field direction (see Fig. 1\textbf{A} in main text), and any perturbation results in an electrostatic torque $\mathcal{T}^{\text{e}} = \mathbf{P}_\mathrm{exp} \times \mathbf{E}$, from the dipole rotation. Nevertheless, even if $\tau_{p} > \tau_{l}$ is satisfied, $\mathcal{T}^{\text{e}}$ needs to overcome the viscous torque exerted on the particle by the liquid to initiate rotation, $\mathcal{T}^{\text{H}} = -\alpha \omega$, where the angular velocity is given by $\omega$ and $\alpha = \pi\eta \sigma^3$ is the rotational friction coefficient. We use polymethyl methacrylate colloids of diameter $\sigma = 2.92 \mu$m, with $\epsilon_{p} = 2.6\epsilon_{0}$, and a 5 mM AOT/hexadecane solution with $\eta = 2.78$ mPA, $s_{l} \approx 10^{-8}\, \Omega^{-1}\,\text{m}^{-1}$ \cite{schmidt2012}, $s_{p} \approx 10^{-14}\,\Omega^{-1}\,\text{m}^{-1}$ \cite{pannacci2007} and $\epsilon_{l} \approx 2\epsilon_{0}$ for our system. The critical threshold is given by

\begin{equation}\label{eqEQ}
E_{{\rm{Q}}}=\frac{1}{2}[\pi\epsilon_{l}\sigma^{3}(\chi^{0}-\chi^{\infty})\tau_{\mathrm{MW}}\alpha^{-1}]^{1/2},
\end{equation}

\noindent
where  the polarisability factors

\begin{equation}\label{eqChi0}
\chi^{0}=\frac{s_{p}-s_{l}}{s_{p}+2s_{l}}
\end{equation}

\noindent
and

\begin{equation}\label{eqInfty}
\chi^{\infty}=\frac{\epsilon_{p}-\epsilon_{l}}{\epsilon_{p}+2\epsilon_{l}}
\end{equation}

\noindent
account for the conductivities and permittivities of the particle and liquid respectively. The characteristic dipole relaxation timescale is given by the Maxwell-Wagner time,

\begin{equation}\label{tauMW}
\tau_{\mathrm{MW}}=\frac{\epsilon_{p}+2\epsilon_{l}}{s_{p}+2s_{l}}.
\end{equation}


\subsection*{Microscopic model of effective interactions in Quincke rollers}

Following Ref. \cite{bricard2013}, we consider a pairwise alignment interaction between rollers
that leads to a torque on particle $i$
\begin{equation*}
\mathcal{T}_i = - \frac{\partial  \mathcal{R}_{\rm align}} {\partial \theta_i} \; ;  \;
\end{equation*}

\begin{widetext}
\begin{equation}
\mathcal{R}_{\rm align}
=
- \sum_{\begin{array}{c} j ,   |\mathbf{r}_{ij}| \le r_{c1} \end{array}} \left( A_1 \hat{\mathbf{P}}_i \cdot \hat{\mathbf{P}}_j + A_2
 (\hat{\mathbf{P}}_i - \hat{\mathbf{P}}_j)\cdot{\hat{\mathbf{r}}}_{ij} +
A_3  \hat{\mathbf{P}}_j \cdot (2\hat{\mathbf{r}}_{ij} \hat{\mathbf{r}}_{ij} - {\mathbf{I}}) \cdot  \hat{\mathbf{P}}_i \right) \;
\label{eqH1}
\end{equation}
\end{widetext}

where $\hat{\mathbf{P}}_i = ( \cos \theta_i, \sin \theta_i )$ is the direction of motion of the $i$th roller,  and ${\bf r}_{ij}$ is the separation between rollers $i$ and $j$. This has the minimum number of terms required to describe the electro-hydrodynamically induced alignment interactions with the correct symmetry and whose range is set by the distance between plates in the experimental setup. We truncate $\mathcal{R}_{\rm align}$ at $r_{c1} = 3.0\,\sigma$, where $\sigma$ is the particle diameter. We note that angular momentum is not conserved by these dynamics.

The electro-osmotic long-ranged attraction \cite{nadal2002} is modelled by a turncated and shifted (at $r_{c2} = 5.0\,\sigma$) potential of the form $\mathcal{H}_{\text {attr}} = -A_4 \exp(-\kappa r)/r^2$, where $\kappa = 1/3\,\sigma^{-1}$ is the inverse screening length. The excluded volume interactions between rollers are represented by a repulsive Weeks-Chandler-Anderson (WCA) interaction of the form $\mathcal{H}_{\text{exc}}=4\epsilon((\sigma/r)^{12}-(\sigma/r)^{6})+\epsilon$, where $\epsilon=k_{\rm B}T$ is the energy unit of the model. The WCA potential is truncated at $r_{c3}=2^{1/6}\sigma$.

The coupling parameters in the alignment interactions are estimated to be
$A_1=0.93k_BT
,~A_2=0.33k_BT
$ and
$~A_3=0.48k_BT$
for our experimental conditions (see SM for more details), and we chose the attraction strength to be $A_4=10k_{\rm B}T$. We verified that the qualitative phase behaviour of the model remains the same if we vary the strength of the long-ranged attraction. We note that we have parametrised $A_1,A_3$ from the single particle dynamics in the dilute gas phase, the attractive interactions $A_2,A_4$ are determined from the experimental parameters. Further details are given in the SM.

\paragraph{Simulation details. ---}
Brownian dynamics simulations were performed on a 2D system composed of $N=10000$ interacting Quincke rollers. We integrate the
over-damped Langevin equations (Eqs. \ref{eqABP}) using the stochastic Euler scheme with a time step of $dt=10^{-5} \tau$. In our simulations, the interparticle force on the $i$th roller $\mathbf{F}_i = -\nabla_i (\mathcal{H}_{\text {attr}} +\mathcal{H}_{\text {exc}})$ while the torque on the $i$th roller $\mathcal{T}_i = -\partial \mathcal{R}_{\rm align} / \partial \theta_i$. The particle diameter $\sigma$, thermal energy $\epsilon=k_{\rm B}T$ and Brownian time $\tau=\sigma^2/D_t$ are chosen as basic units for length, energy and time, respectively. We take 
$D_r=3D_t/\sigma^2$, as expected for an spherical particle in the low-Reynolds-number regime. We study the phase behaviour of the system as a function of two dimensionless parameters; P\'eclet number $\text{$\pecl$} = f^p \sigma / k_{\text{B}}T$ and the area fraction $\ds \phi=\frac{N\pi\sigma^2}{4L^2}$, where $L$ is the linear size of the simulation box.

\paragraph{Order parameter details. ---}
Here we take the mean of the bond-orientational order parameter $\psi_6$ across $N$ particles

\begin{equation}
\label{eqPsi6Mean}
\psi_6 =\frac{1}{N}\sum_{j=1}^{N}|\psi^{j}_{6}|.
\end{equation}

The value of the order parameter for each particle is

\begin{equation}
\label{eqPsi6}
\psi_{6}^j \equiv \frac{1}{Z_{j}}\sum_{k=1}^{Z_{j}}\exp\Big(i6\theta_{k}^{j}\Big)
\end{equation}

\noindent
where $Z_{j}$ is the co-ordination number of particle $j$ obtained from a Voronoi construction and $\theta_{k}^{j}$ is the angle made between a reference axis and the bond between particle $j$ and its $k$th neighbour. $\psi_{6}=1$ indicates perfect hexagonal ordering, whereas completely disordered structures give $\psi_{6}=0$. Fig.\,\ref{figStructure}\textbf{A} shows that for a passive Brownian system there is almost no hexatic order.

We quantify the fluctuations in $\psi_6$ by defining the susceptibility

\begin{equation}
\label{eqChi6}
\chi_{6} \equiv \langle \psi_6^2 \rangle - \langle\psi_6\rangle^2
\end{equation}
\noindent
where $\psi_6^2  = 1/N \sum_{j=1}^{N} |\psi_6^j|^2$.
\\

\subsection*{Supplementary Materials}

\begin{figure*}
  \centering
  \includegraphics[width=0.35\textwidth]{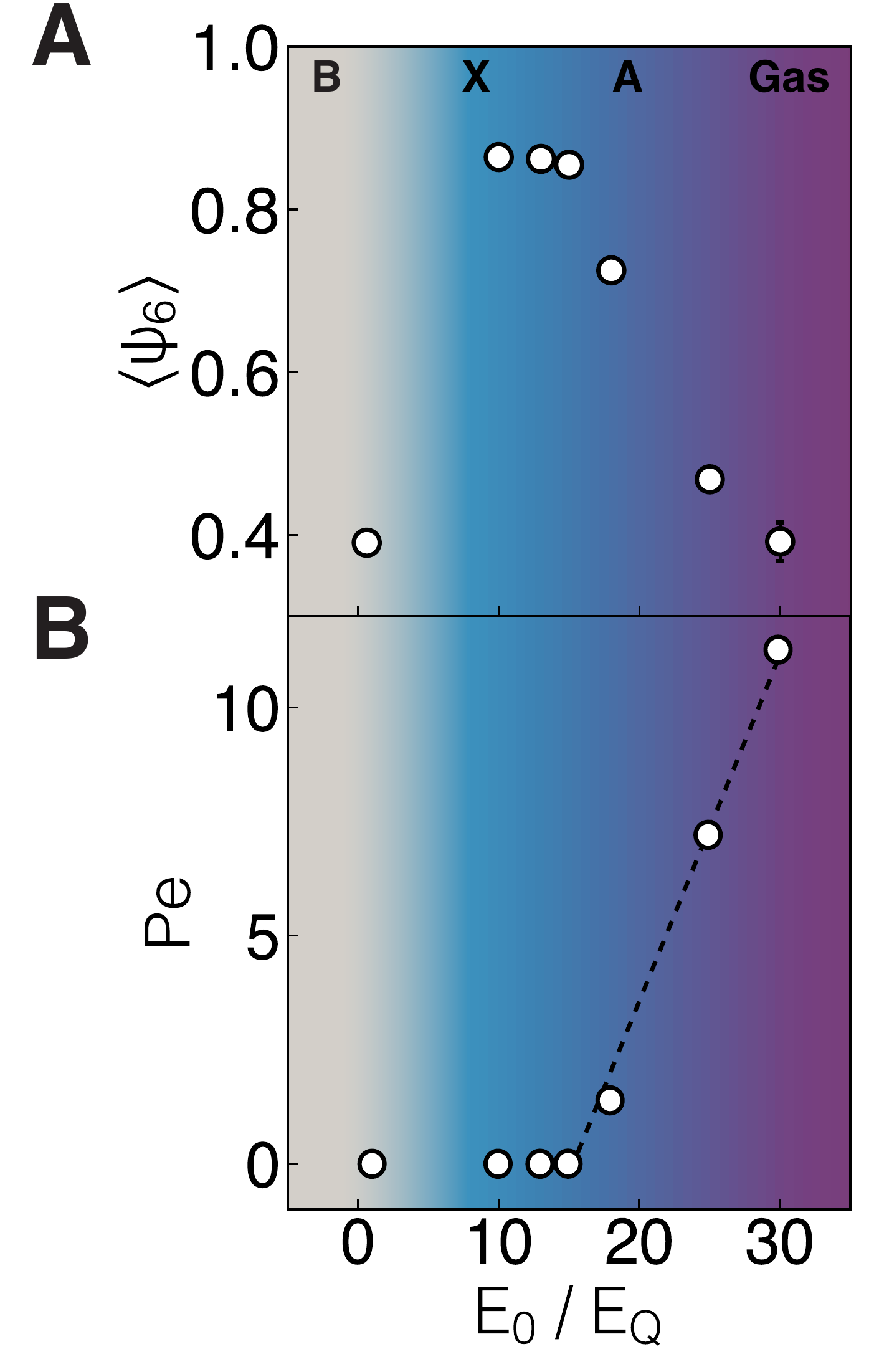}
  \caption{
  \textbf{a.} Mean local hexagonal order parameter.
  \textbf{b.} P\'eclet number as a function of field strength deduced form experimental measurement at low area fraction (see main text for details).
  As in the main text, B denotes Brownian fluid (i.e. low field stength), X (passive) crystals, A amoeba and gas is the active gas.
  }
\label{figBoopPeField}
\end{figure*}

\begin{figure*}
\centering
\includegraphics[width=0.95\textwidth]{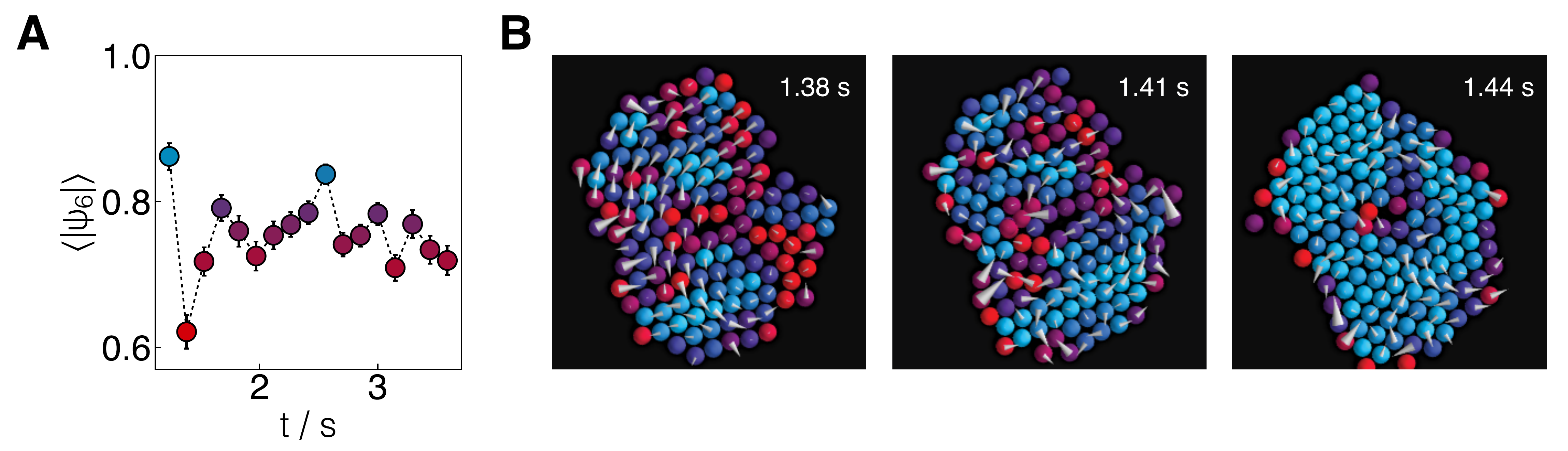}
\caption{\textbf{a.} Mean hexagonal order parameter in time for amoeba aggregates. \textbf{b.} Frame time sequence from Fig. 1c in the main text . Colours on particles indicate the local hexagonal order, $\psi^{i}_{6}$ (see main text for details)}.
\label{sFigAmoebaBoop}
\end{figure*}


\begin{figure*}
\centering
\includegraphics[width=0.85\textwidth]{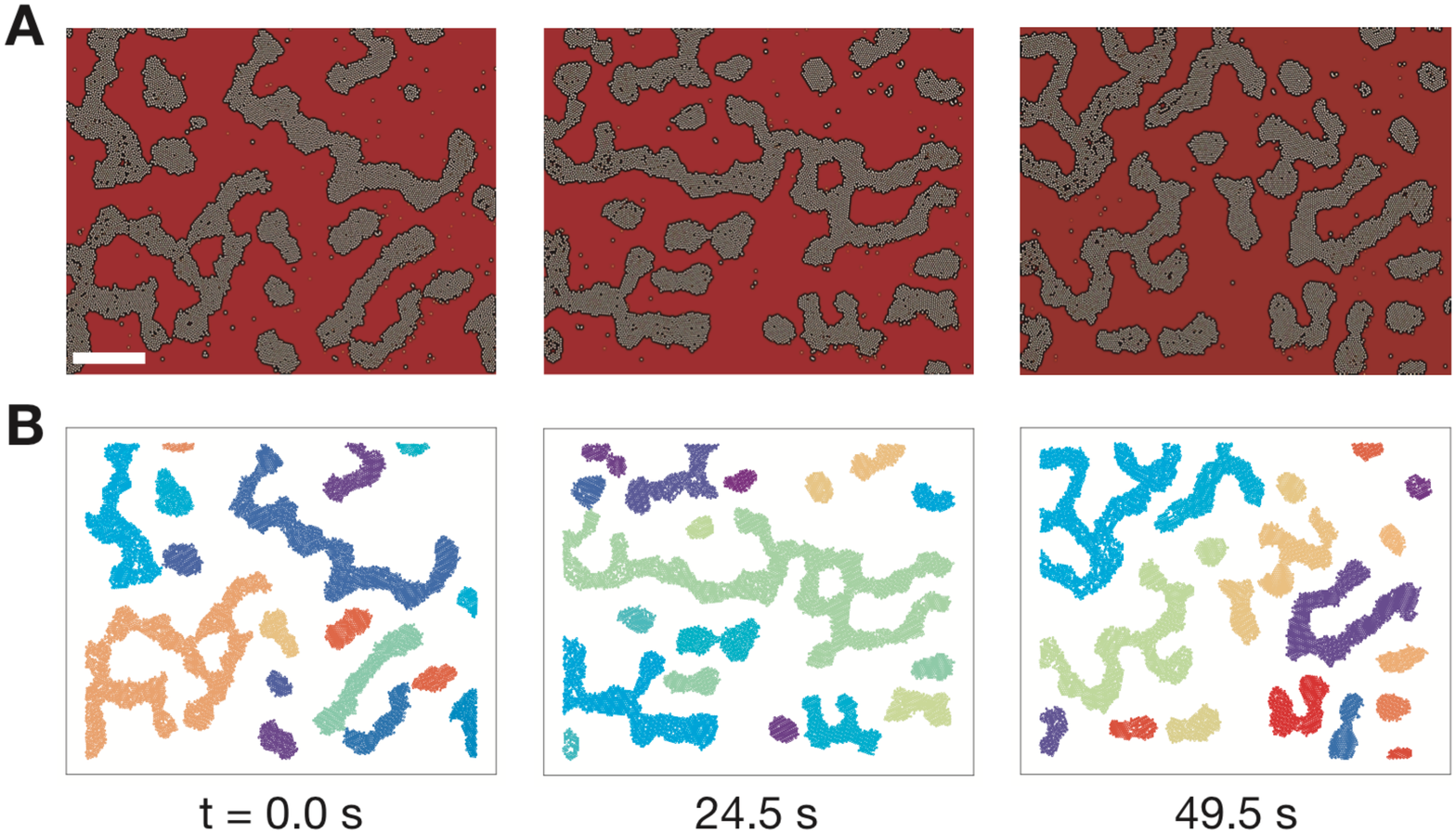}
\caption{\textbf{a.} Image sequence of amoeba clusters. Scale bar is $100\,\mu\rm{m}$. \textbf{b.} Single cluster are shown in different colours.}
\label{sFigAmoebaClusters}
\end{figure*}


\section*{Microscopic model of Alignment Interactions in Quincke Rollers}

The following description is based on a microscopic model describing the dynamics of a population of colloidal rollers due to Quincke rotation. The direct interactions
are detailed in the Methods, and are captured in the force $\mathbf{F}_i$ in Eq. \ref{eqTranslation}. Here we consider the alignment terms. The equations of motion for the $i$th self-propelled particle are given by the following Langevin equations, where for the rotational case we have rewritten the version in the main text to explicitly consider the effective alignment interaction.

\begin{equation}\label{eqTranslation}
\dot{\mathbf{r}}_i = \frac{D_t}{k_BT} [ {\mathbf{F}}_i + f^p \hat{\mathbf{P}}_i ] + {\sqrt{2D_t}} \bm{\xi}_i^{t}
\end{equation}

\noindent
and
\begin{widetext}
\begin{equation}\label{eqRotation}
	\dot{\theta_i}=-\frac{D_r}{k_BT}\frac{\partial}{\partial\theta_{i}}\sum_{j\not=i}\mathcal{R}_{\rm{align}}({\mathbf{r}_{ij}},{\mathbf{\hat{P}}}_{i},{\mathbf{\hat{P}}}_{j})+\sqrt{2D_{r}}\xi_{i}^r
\end{equation}
\end{widetext}

\noindent
where the particle $i$ is subject to a propulsion force of magnitude $f^p$ whose direction changes due to the alignment interaction and noise $\xi_{i}$. Note that because the simulations are strictly in 2D, the direction of the dipole $\mathbf{P}$ in Eq. \ref{eqRotation} is that of the rotation, \emph{i.e.} the direction of self-propulsion, rather than the (3D) induced dipole of the experimental system $\mathbf{P}_\mathrm{exp}$ mentioned above.

Introduced by Caussin and Bartolo \cite{bricard2013}, the effective alignment interaction $\mathcal{R}_{\rm{align}}$ reads

\begin{widetext}
\begin{equation}\label{potential}
\mathcal{R}_{\rm{align}}({\mathbf{r}},{\mathbf{\hat{P}}}_{i},{\mathbf{\hat{P}}}_{j})=-A_1(r){\mathbf{\hat{P}}}_{i}\cdot{\mathbf{\hat{P}}}_{j}-A_2(r)\mathbf{\hat{r}}\cdot(\mathbf{\hat{P}}_{i}-\mathbf{\hat{P}}_{j})-A_3(r){\mathbf{\hat{P}}}_{j}\cdot(2{\mathbf{\hat{r}\hat{r}-I}})\cdot{\mathbf{\hat{P}}}_{i}
\end{equation}
\end{widetext}

\noindent
having ${\mathbf{\hat{r}}}\equiv{\mathbf{r}}/r$. The coefficients $A_1(r), A_2(r)$ and $A_3(r)$ incorporates the microscopic parameters and are given by:

\begin{subequations}
\begin{widetext}
\begin{equation}\label{eqA1}
\renewcommand{\theequation}{\theparentequation.
\arabic{equation}}
A_1(r)=3\tilde{\mu}_{s}\frac{\sigma^3}{8r^3}\Theta(r)+9\Bigg(\frac{\mu_{\perp}}{\mu_{r}}-1\Bigg)\Bigg(\chi^{\infty}+\frac{1}{2}\Bigg)\Bigg(1-\frac{E_{{\rm{Q}}}^{2}}{E_0^2}\Bigg)\frac{\sigma^{5}}{32r^5}\Theta(r)
\end{equation}
\end{widetext}

\noindent
accounting for the short-ranged hydrodynamic interactions and electrostatic couplings that promote the alignment of directions between particles $i$ and $j$. Here, $\mu_{\perp}$ and $\mu_{r}$ are the mobility coefficients depending on the liquid viscosity and the distance $d$ between the surface and particle respectively. From the expressions in \cite{ONeill1967,Goldman1967a,Goldman1967,LIU2010} we obtain $\chi^{\infty}=0.08$, $\tilde{\mu}_{s}=11$ and $\mu_{\perp}/\mu_{r}=1.5$.\\

The electrostatic repulsion and the electro-hydrodynamic interactions coupling are encoded in the $A_2(r)$ and $A_3(r)$ coefficients respectively,

\begin{widetext}
\begin{equation}\label{eqA2}
\renewcommand{\theequation}{\theparentequation.
\arabic{equation}}
A_2(r)=6\Bigg(\frac{\mu_{\perp}}{\mu_{r}}-1\Bigg)\sqrt{\frac{E_0^{2}}{E_{\rm{Q}}^{2}}-1}\Bigg[\Bigg(\chi^{\infty}+\frac{1}{2}\Bigg)\frac{E_0^{2}}{E_{\rm{Q}}^{2}}-\chi^{\infty}\Bigg]\frac{\sigma^{4}}{16r^4}\Theta(r)
\end{equation}

\begin{equation}\label{eqA3}
\renewcommand{\theequation}{\theparentequation.\arabic{equation}}
A_3(r)=2\tilde{\mu_{s}}\frac{\sigma^{2}}{4r^{2}}\frac{\sigma}{2H} + \Bigg[\tilde{\mu_{s}}\frac{\sigma^{3}}{8r^{3}} + 5\Bigg(\frac{\mu_{\perp}}{\mu_{r}}-1\Bigg)\Bigg(\chi^{\infty}+\frac{1}{2}\Bigg)\Bigg(1-\frac{E_{{\rm{Q}}}^{2}}{E_0^2}\Bigg)\frac{\sigma^{5}}{32r^{5}}\Bigg]\Theta(r)
\end{equation}
\end{widetext}

\noindent
where the hydrodynamic and electrostatic couplings are screened over distances proportional to the chamber distance, H = 16.2 $\mu$m. A more detailed description can be found in Refs. \cite{bricard2013}, and  \cite{bricard2015}. We estimate such coefficients considering the experimental field intensity under which we observe the active gas phase ($1.85 \times 10^{6}V $m$^{-1}$), and average them over distances $r\in[\sigma,3\sigma]$. For convenience we approximate the screening function as $\Theta(r)=1$ if $r \leq H/\pi $ and $\Theta(r)=0$ otherwise.

Under these assumptions, we obtain
\begin{equation}\label{A3}
\renewcommand{\theequation}{\theparentequation.\arabic{equation}}
	A_1=0.93 k_BT
\end{equation}
\begin{equation}\label{B3}
\renewcommand{\theequation}{\theparentequation.\arabic{equation}}
	A_2=0.33 k_BT
\end{equation}
\begin{equation}\label{C3}
\renewcommand{\theequation}{\theparentequation.\arabic{equation}}
	A_3=0.48 k_BT
\end{equation}
\end{subequations}

\subsection{Supplementary Movies}

\section{Supplementary Movie 1}

\textit{Amoeba aggregates ---}  Finite-size amoeba clusters displaying collective rotation. The interaction between amoeba aggregates leads to merging. Movie is coloured with a glow effect for clarity. Colloid diameter $\sigma = 2.92\,\mu m$. Field strength, $E_0 = 19.4 E_{Q}$, and Pe = 1.5. Frame acquisition at 100 fps, movie played at 34 fps \cite{movies}.

\section{Supplementary Movie 2}

\textit{Amoeba phase ---} Movie shows the experimental trajectory of an amoeba-like aggregate. The colourbar indicates the local hexagonal order parameter $\psi^{i}_{6}$ for each particle. Fluctuations of the order parameter result as the aggregates merge and break. White arrows indicate the instantaneous collective displacement. Colloids diameter $\sigma = 2.92\,\mu m$. Field strength, $E_0 = 19.4 E_{Q}$, and Pe = 1.5. Frame acquisition at 100 fps, movie played at 17 fps \cite{movies}.

\section{Supplementary Movie 3}

\textit{Experimental phase transition ---} Transition of an isolated cluster with the increase on $E_0$. The transition goes from a highly ordered and dynamically arrested state to an isotropic state of Quincke rollers. The colourbar indicates the local hexagonal order parameter $\psi^{i}_{6}$. Particles in blue posses high hexagonal order, whereas the order is poor in particles in red. Colloids of diameter $\sigma = 2.92\,\mu m$ Field strength $E_0\,\in [9.9,29.8]E_{Q}$, and Pe $\in[10^{-4},11.3]$. Frame acquisition at 100 fps, movie played at 17 fps \cite{movies}.

\section{Supplementary Movie 4}

\textit{Onset of banding  ---} Thick bands (of tens of particles) form as particle trajectories undergo alignment. The bands propagate through an isotropic state. Pe = 50 \cite{movies}.

\section{Supplementary Movie 5}

\textit{Onset of banding  ---}
Same data as Supplementary Movie 4, zooming out sequence.
Pe = 50 \cite{movies}.
\\
\subsection*{Acknowledgements}
The authors would like to thank Denis Bartolo, Olivier Dauchot, Jens Eggers, Mike Hagan, Rob Jack, Cristina Marchetti, Sriram Ramaswamy, Thomas Speck and Chantal Valeriani for helpful discussions. CPR, JH and FT would like to acknowledge the European Research Council under the FP7 / ERC Grant agreement n$^\circ$ 617266 ``NANOPRS''. AMA is funded by CONACyT. TBL and MM are supported by BrisSynBio, a BBSRC/EPSRC Advanced Synthetic Biology Research Center (grant number BB/L01386X/1). Part of this work was carried out using the computational facilities of the Advanced Computing Research Centre, University of Bristol.

\subsection{References and Notes}

\bibliographystyle{naturemag}
\bibliography{amoeba}

\end{document}